\documentclass{icrc2009}

\usepackage{graphicx}   
\usepackage[caption=false]{caption}    
\usepackage[font=footnotesize]{subfig} 
\usepackage{fixltx2e}
\usepackage{url}

\newcommand{\shorttitle}[1]%
{\markboth{Proceedings of the 31\MakeLowercase{$^{st}$} ICRC, {\L}\'{o}d\'{z} 2009}{#1} }
\newcommand{\etal}{\MakeLowercase{\textit{et al. }}} 

\begin{document}
\title{The June 2008 flare of Markarian 421 from optical to TeV energies}

\author{\IEEEauthorblockN{
R. M. Wagner\IEEEauthorrefmark{1}, 
I. Donnarumma\IEEEauthorrefmark{2},
J. Grube\IEEEauthorrefmark{3}, 
M. Villata\IEEEauthorrefmark{4},
C. M. Raiteri\IEEEauthorrefmark{4}, 
C.-C. Hsu\IEEEauthorrefmark{1}, \\
K. Satalecka\IEEEauthorrefmark{5}, 
E.Bernardini\IEEEauthorrefmark{5} and
P. Majumdar\IEEEauthorrefmark{5} \\ 
for the MAGIC, VERITAS, AGILE, GASP-WEBT collaborations}
                            \\
\IEEEauthorblockA{\IEEEauthorrefmark{1}Max-Planck-Institut f\"ur Physik, D-80805 Munich, Germany}
\IEEEauthorblockA{\IEEEauthorrefmark{2}INAF/IASF-Roma, I-00133 Roma, Italy}
\IEEEauthorblockA{\IEEEauthorrefmark{3}School of Physics, University College Dublin, Dublin 4, Ireland}
\IEEEauthorblockA{\IEEEauthorrefmark{4}INAF, Osservatorio Astronomico di Torino, Italy}
\IEEEauthorblockA{\IEEEauthorrefmark{5}Deutsches Elektronen-Synchrotron, D-15738 Zeuthen, Germany}
}

\shorttitle{Wagner \etal The June 2008 flare of Mkn 421}
\maketitle

\begin{abstract}
We present optical to very-high energy (VHE) gamma-ray observations of Mrk 421
between 2008 May 24 and June 23. A high-energy (HE) gamma-ray signal was
detected by AGILE-GRID during June 9--15, brighter than the average flux
observed by EGRET in Mrk 421 by a factor of $\approx$ 1.5. In 20--60 keV X-rays,
a large-amplitude 5-day flare (June 9--15) was resolved with a maximum flux of
approx. 55 mCrab. SuperAGILE, RXTE/ASM and Swift/BAT data show a clearly
correlated flaring structure between soft and hard X-rays, with a high
flux/amplitude variability in hard X-rays. Hints of the same flaring behavior
is also detected in the simultaneously recorded GASP-WEBT optical data. A target of
opportunity observation by Swift near the flare maximum on June 12--13 revealed
the highest 2-10 keV flux ever observed ($>$100 mCrab) and a peak synchrotron
energy of approx. 3 keV, a large shift from typical values of 0.5--1 keV.
Observations at VHE ($E>$200 GeV) gamma-rays during June 6--8 show the source
flux peaking in a bright state, well correlated with the simultaneous peak in
the X-rays.  The gamma-ray flare can be interpreted within the framework of the
Synchrotron Self Compton model in terms of a rapid acceleration of leptons in
the jet.
\end{abstract}

\begin{IEEEkeywords}
 X-rays: observations; Gamma-rays: observations; Blazars: individual (Mkn 421)
\end{IEEEkeywords}

\section{Introduction}
Mrk 421 is a nearby blazar ($z=0.031$), detected in $\gamma$-rays by EGRET
\cite{lin92} and it was the first extragalactic object detected at $E> 500$ GeV
\cite{pun92}. It belongs to the class of high-energy peaked BL Lac objects
(HBLs) \cite{pad95}, i. e. radio-loud active galactic nuclei with high radio
and optical polarization. Its spectral energy distribution (SED) is
double-humped with a first peak usually in the soft to medium X-ray range, and
a second one at GeV-TeV energies \cite{sam96,fos98}.  The first hump is
commonly interpreted as due to synchrotron radiation from high-energy electrons
in a relativistic jet, while the origin of the second peak is still uncertain.
In leptonic scenarios it is interpreted as inverse Compton (IC) scattering of
the synchrotron (Synchrotron self-Compton, SSC) or external photons (External
Compton, EC)  by the same population of relativistic electrons. The observed
correlated variability between X-rays and TeV $\gamma$-rays
\cite{mar99,fos08,wag08} is well explained in the SSC framework \cite{ghi98},
whereas the EC scenario is unlikely to apply in HBLs, due to the low density of
ambient photons.  Alternatively, hadronic models invoke proton-initiated
cascades and/or proton-synchrotron emission \cite{aha00,mue03}.  Leptonic and
hadronic scenarios for HBLs predict different properties of the $\gamma$-ray
emission in relation to emissions in other energy bands.  $\gamma$-ray
observations of flaring BL Lacs and simultaneous multiwavelength data are thus
the keys to investigating these two scenarios.

A hard X-ray flare of Mrk 421 was detected by SuperAGILE on 2008 June 10
\cite{cos08}. This detection was later followed by a detection in $\gamma$-rays
\cite{pit08} by the AGILE/GRID (Gamma Ray Imaging Detector) and prompted a ToO
observation by Swift/XRT, complementing the ongoing multifrequency observing
campaign of Mrk 421 with GASP-WEBT (optical), MAGIC and VERITAS (TeV).  A full
account of this work has recently been published \cite{paper}.

\section{Observations and Results}
\subsection{AGILE observations}
The AGILE \cite{tav08} composite payload allows for simultaneous observations
in the energy ranges 30 MeV-30 GeV and 20--60 keV over a very wide field of view
by means of GRID and the hard X-ray imager SuperAGILE, respectively. Mrk 421
was observed for five days, between 2008-06-09 17:02 UT and 2008-06-15 02:17
UT.

\subsubsection{Hard X-ray observations}
On 2008 June 10, SuperAGILE detected enhanced hard X-ray emission from Mrk 421.
The measured flux in 20--60 keV was found to be above 30 mCrab (24-hour
average), almost an order of magnitude larger than its typical flux in
quiescence.  In the following days, the flux increased up to about 55 mCrab.
The 5-day 20-60 keV SuperAGILE light curve is shown in Fig. 1c.  Using the
publicly available light curves for this source from the BAT instrument in the
15-50 keV energy range, we calculated daily averages, obtaining a good coverage
also before and after the AGILE observation (black squares in Fig. 1c),
revealing that SuperAGILE indeed observed the maximum brightness of this hard
X-ray flare. 

SuperAGILE photon-by-photon data allows extraction of a time-averaged energy
spectrum from the mask-convolved data. Given the lack of substantial spectral
variability in the SuperAGILE/ASM hardness ratio (Fig. 1e) we accumulated the
average energy spectrum from the data of the last 4 days of the observations,
when the source flux varied between 35 and 55 mCrab, for a total exposure of
$\sim 140$ ks, constraining the photon index of a simple
power law, $\Gamma =2.43^{+0.69}_{-0.64}$
($\chi^{2}_{\rm dof}=0.8$, 2 dof). The average flux is $F(20-60 \mathrm {keV})
=(4.90 \pm 0.54) \times
10^{-10}$ erg cm$^{-2} \rm s^{-1}$ (($9.8 \pm 1.1) \times
10^{-3}$ photons cm$^{-2}$
s$^{-1}$).

\subsubsection{Gamma-ray Observations}

Mrk 421 was not detected on daily time scales, implying a daily-averaged flux
$<100 \times 10^{-8}$ photons cm$^{-2}$ s$^{-1}$, similar to what observed by
EGRET \cite{har99}. A 4.5-$\sigma$ significance in the range 100 MeV-10 GeV
resulted from an integration over
the whole 5-day period ($\sim 260$ ks). The measured flux is  ($42^{+14}_{-12})
\times 10^{-8}$ photons cm$^{-2}$ s$^{-1}$, about $\sim$3 times
higher than the average flux detected by EGRET ($\sim13 \times
10^{-8}$, \cite{har99}) and $\sim$1.5 times higher than, but still
consistent, with the highest flux ($27\pm 7)\times 10^{-8}$ photons cm$^{-2}$
s$^{-1}$ observed during the Viewing Period 326. 

\subsection{The Soft X-ray band}

\subsubsection{Swift X-ray telescope}

Following the SuperAGILE detection, on 2008 June 12 Swift 
ToO observations were triggered.
between 2008 June 12
19:33:20 UT and June 13 at 01:57:37 UT 
 for 5 ks.
The XRT spectral data are well
described by an absorbed log-parabolic model. A joint fit of the XRT 
and SuperAGILE spectral data using the 4-day average spectrum 
used the log-parabolic model with Galactic absorption ($N_{\rm H}^{\rm Gal} = 1.61 \times
10^{20}$\,cm$^{-2}$; \cite{loc95}), defined as:
\begin{center}
$F(E) = K E^{-a-b \log(E)}$  photons cm$^{-2}$ s$^{-1}$
\end{center}
where $a$ is the photon index at 1 keV and $b$ is the curvature
parameter \cite{mas04,mas08}. This
model usually describes adequately the featureless and curved
spectrum in HBLs.
The joint fit provides
$a=1.65_{-0.02}^{+0.01}$, $b=0.37_{-0.005}^{+0.01}$
($\chi^{2}_{\rm dof}$=1.4, 763 dof),  which implies  a peak energy
$2.97^{+0.22}_{-0.18}$ keV, and predicts $F_{\rm 2-10keV} =
2.56\times10^{-9}$\,erg\,cm$^{-2}$\,s$^{-1}$ (or $0.4$
photons\,cm$^{-2}$\,s$^{-1}$) and $F_{\rm 20-60keV}=
(5.7\pm 0.6)\times10^{-10}$\,erg\,cm$^{-2}$\,s$^{-1}$ (($1.1 \pm 0.1) \times 10^{-2}$
photons\,cm$^{-2}$\,s$^{-1}$), comparable to the
stand-alone SuperAGILE best fit.

\subsubsection{RossiXTE All Sky Monitor (ASM)}

We retrieved the public
light curves provided by the
ASM to trace the evolution of
the soft X-rays during the AGILE observation.
Fig. 1b shows the daily light curve of Mrk 421
in the energy range 2-12 keV, obtained by properly averaging the
dwell-by-dwell data.

The emission at soft X-rays is well correlated with hard
X-ray emission. The ASM data show that the XRT observation took
place when the source was at its maximum emission at soft X-rays
(MJD $\sim$ 54630). Comparing the relative intensity
of the two flares in Fig. 1, the second peak appears to be
significantly harder than the first one. This is also seen in Fig. 1e, 
where we computed the daily-averaged hardness ratio between hard (15--60 keV) 
and soft (2--12 keV) X-rays. The source appears to have undergone the hardest 
part of this double-humped flare just during the AGILE $\gamma$-ray detection.

\begin{figure}
\begin{center}
\includegraphics[width=0.48\textwidth]{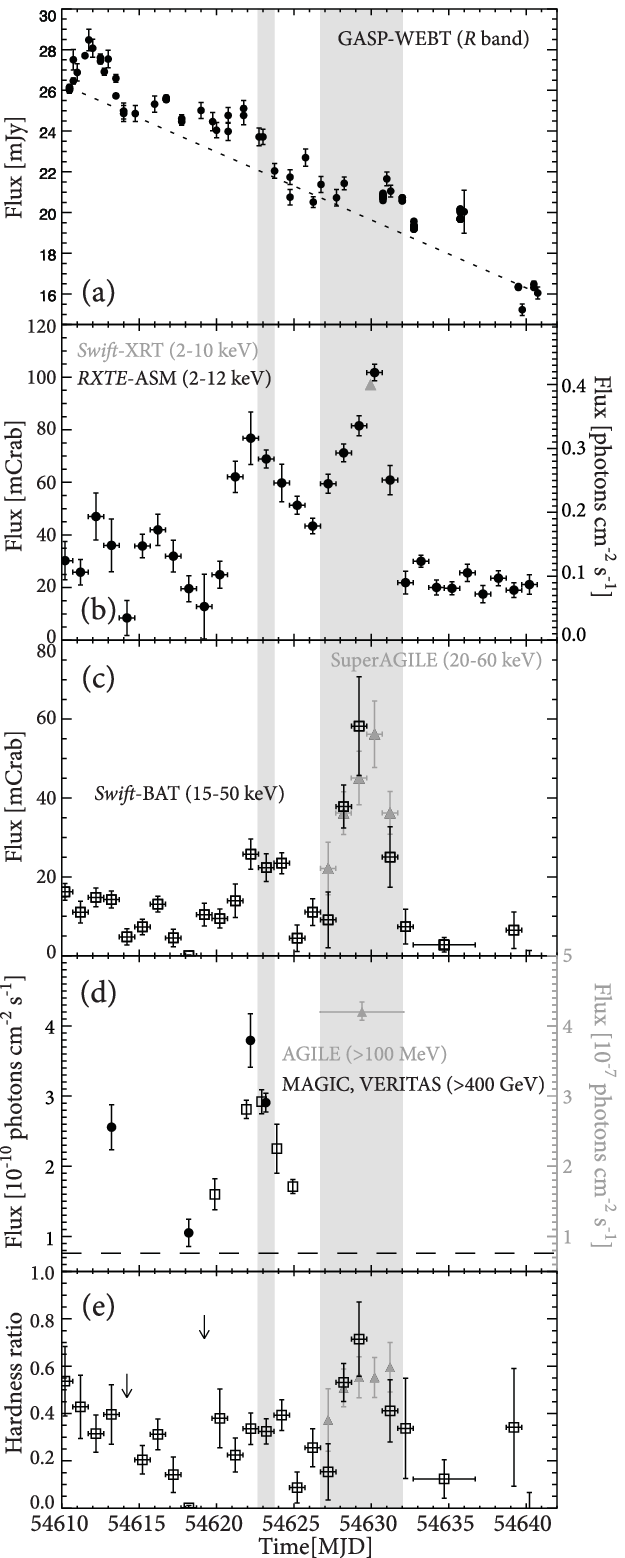}
\vspace*{-.4cm}
\end{center}
 \caption{\footnotesize{\textbf{a)} $R$-band optical light curve from
GASP-WEBT (May 24--June 23); \textbf{b)} ASM (2-12 keV) light
   curve (bin size is 1 day) and XRT (2-10 keV) flux (triangle);
   \textbf{c)} SuperAGILE (20-60 keV, triangles; 1 Crab $= 0.2 $ ph cm$^{-2}$ s$^{-1}$)
   and BAT (15-50 keV, empty squares; 1 Crab $= 0.29$ ph cm$^{-2}$
   s$^{-1}$); \textbf{d)} MAGIC and VERITAS ($>$400 GeV,
   empty squares and filled circles, respectively), the Crab
   flux $> 400$ GeV (horizontal dashed line), AGILE ($> 100$ MeV, triangle and right vertical axis); \textbf{e)} the hardness ratio computed by using the
  SuperAGILE and ASM data for each day. The shaded areas mark
  period 1 and period 2.}}
\end{figure}

\subsection{Observations in the  VHE $\gamma$-rays}

The 4-telescope array VERITAS \cite{acc08} 
in Arizona, USA, and the single-dish instrument MAGIC \cite{bai04,tes07} 
at La Palma are imaging air Cherenkov telescopes covering an energy range 
from $\sim 100$ GeV to some tens of TeV. The instruments have a typical 
energy resolution of $<$20\% (VERITAS) and 20--30\% (MAGIC), and 
event-by-event angular resolution of $<$ 0.14$^\circ$. Wobble-mode 
observations \cite{dau97} at an 0.4$^\circ$ offset from the camera 
center were taken on five nights from 2008 June 3--8 with MAGIC at zenith 
angles (ZA) between 28$^\circ$ and 48$^\circ$ and on 4 nights (May 27, 
June 1, 5, 6) with VERITAS (wobble offset: 0.5$^\circ$) at ZA between 
32$^\circ$ and 40$^\circ$ during partial moon light conditions. The total 
live-time after applying quality selection is 2.95 and 1.17 hours with 
MAGIC and VERITAS, respectively. The data were analyzed using the MAGIC 
\cite{alb08} and VERITAS \cite{dan07,acc08} 
standard calibration and analyses and image parameters \cite{hil85}. 
Two independent analyses of both the MAGIC and VERITAS data sets yielded 
consistent results. In total, a signal corresponding to a significance level 
of $44\sigma$ (VERITAS) and $66\sigma$ (MAGIC) is obtained.
The combined MAGIC-VERITAS data (Fig. 1d) show a 
transient peaking near MJD 54622. The VERITAS 
energy spectrum for June 6 is provided. A power law fit over the energy range 
0.3--5 TeV resulted in a $\chi^{2}_{\rm dof}$=0.7 with a photon index 
$\Gamma = 2.78 \pm 0.09$. In Fig. 2 we show the intrinsic $\gamma$-ray spectrum at the source
reconstructed by removing attenuation effects by the extragalactic 
background light \cite{hau01} following the procedure of \cite{rau08}. 
Fitting a power law to the intrinsic spectrum yields a photon 
index $\Gamma = 2.59 \pm 0.08$, which is not significantly harder than the 
measured spectrum due to the relatively low redshift z $=$ 0.031.

\subsection{Optical and UV observations}

Mrk 421 is one of the 28 $\gamma$-ray-loud blazars 
regularly monitored by the GLAST-AGILE Support Program
(GASP; \cite{vil08}) of the Whole Earth Blazar Telescope
(WEBT). The GASP observations
started in early 2007 November. 

During the Swift pointing on 2008 June 12--13, the UVOT \cite{poo08}
instrument observed Mrk 421 in the UVW1 and UVW2 photometric bands. The
UVOTSOURCE tool is used to extract counts, correct for coincidence losses,
apply background subtraction, and calculate the source flux.

\section{Discussion}

Mrk 421 showed an interesting broad-band activity
during the first half of 2008 June as derived from AGILE data
combined with those of GASP-WEBT, Swift, MAGIC and VERITAS.
Using our multi-frequency data we were able to derive
time-resolved SEDs (Fig. 2). We distinguish two time periods: {\it period 1}, 2008 June
6, and {\it period 2}, 2008 June 9--15.
The optical, soft and hard X-ray bands strongly constrain the SED around 
the synchrotron peak, and its daily variability reveals the
physical processes of Mrk~421. Possible correlated variability is shown in
Fig.~1 between the optical, the X-ray, and the high-energy parts of the
spectrum. Based on the physical constraints obtained for the
synchrotron peak, we can model both the HE and VHE $\gamma$-ray 
emission. 
We first model the synchrotron peak of emission using the period 1
optical, soft and hard X-ray data. The short
time-variability (Fig.~1) constrains the size of the emitting region to
$R<cT\delta\sim 5\times 10^{16}(\delta/20)$ cm. Hence,  we consider a one-zone SSC
model \cite{tav98} based on a blob of comoving size
$R=4\times 10^{16}$ cm, with a relativistic Doppler factor $\delta=20$  and
characterized by non-thermal comoving electron energy
distribution function described by a double power law:
\begin{equation}
n_{e}(\gamma)=\frac{K\gamma_{b}^{-1}}{(\gamma
/\gamma_{b})^{p_1}+(\gamma /\gamma_{b})^{p_2}}
\end{equation}
where the comoving Lorentz factor ($\gamma$) varies in the range  $\gamma_{min}=4 \times 10^{3}<\gamma<\gamma_{max}=1.3\times 10^{6}$, the normalization (density)
constant  $K=4\times 10^{-4}$ cm$^{-3}$, and the break energy $\gamma_{b}=
3.6\times 10^{5} $ and
with $p_1=2.22$, $p_2=4.5$, the low-energy and high-energy power-law
indices, respectively (see Table 1).  With these parameters we found that the data
for period 1  are best fitted with a comoving magnetic field $B=0.1$
G.

Variability may be caused by several factors; we consider two
cases: (A) hardening/softening of the electron energy
distribution function caused by particle acceleration processes;
(B) increase/decrease of the comoving particle density,
as a consequence of additional particle injection/loss by shock
processes.

We expect TeV variability to be comparable with the X-ray one if case
(A) applies: this is because the emission is in the Klein-Nishina
regime. Alternately, for case (B) we expect the TeV relative variability
($\Delta F/F$) to be a
factor of 2 greater than that of the X-ray flux variability.

Our AGILE, MAGIC and VERITAS data appear to support case (A). We
compare the SEDs for period 1 and period 2, to
better assess the spectral evolution.
In Fig.~2 we show our optimized modelling of the time-resolved synchrotron peak and consequent SSC high-energy emission for the period 1 as well as for 2008 June 12--13 of the period 2.
In the last case, the adopted model parameters are $p_1=2.1$, $p_2=5$,
$\gamma_{b}=4.2\times 10^{5}$, $K=6\times 10^{-4}$ cm$^{-3}$. 
Our model predicts an even
larger TeV flux for period 2 (no TeV observations exist, however) than
detected in period 1. 

The optical light curve shows 
variations of the order of $10\%$ on a time scale $\sim$few 
days, superimposed on a long decay during the entire period. 
Individual soft and hard X-ray peaks result in increased fluxes by a 
factor of $\sim$2.5 and $\sim$5, respectively: no long term decay appears. This different behavior of 
the X-ray radiation and the bulk of the optical emission may 
suggest a more complex scenarios than A) and B): optical and X 
radiation comes from two different jet regions, each one characterized 
by its own variability. A possible scenario is one in which the inner 
jet region would produce the X-rays and it would be at least partially 
transparent to the optical radiation. 
In contrast, the outer region can only produce lower-frequency 
emission. The signature of the X-ray events visible in the optical 
light curve would come from the inner region and would be diluted 
by the optical radiation emitted from the outer region, see \cite{vil99} 
for the case of Mrk 501; \cite{vil04}. 

Interestingly, the 2-10 keV flux measured by XRT on June 12 --
13,  $\sim 2.6\times 10^{-9}$ erg cm$^{-2}$ s$^{-1}$, is higher
than all previous observations ($< 2 \times 10^{-9}$ erg cm$^{-2}$
s$^{-1}$; \cite{fos08,lic08}). A joint
analysis of the XRT and SuperAGILE data, covering the range from
0.7 to 60 keV, provides a best-fit spectral model consistent with
a log-parabolic shape, with parameters implying a peak energy $\sim$ 3 keV, in good agreement with the steeper
positive correlation between the peak energy and the maximum of
the SED found by \cite{tra07}. 
although our value of the peak energy shows a significant shift
with respect to typical values of 0.5-1 keV for this source.

\begin{figure}
\begin{center}
\includegraphics[width=.42\textwidth]{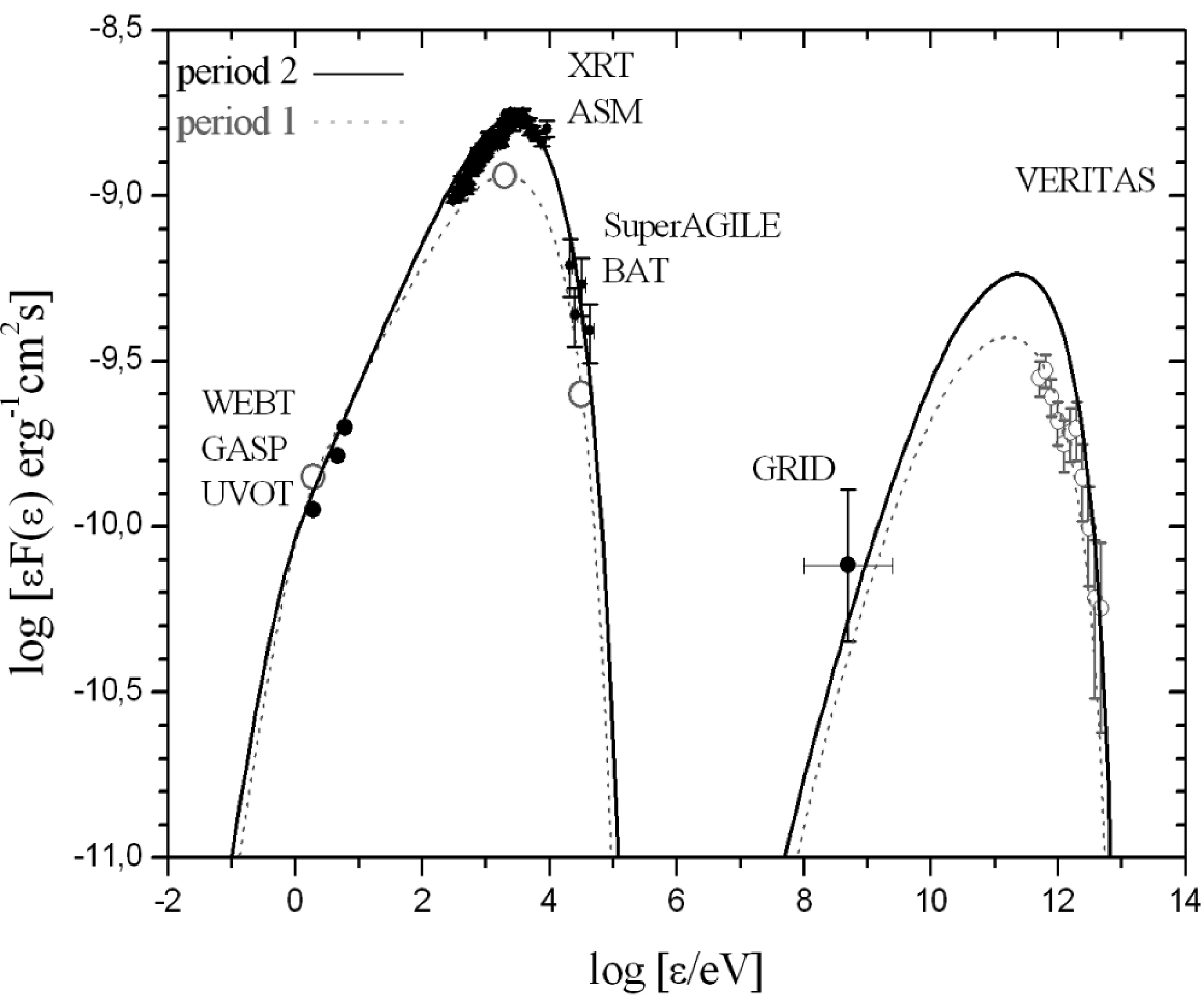}
\caption{SEDs of Mrk 421
obtained by combining the GASP-WEBT, SWIFT/UVOT, RossiXTE/ASM, XRT,
SuperAGILE, BAT, GRID and VERITAS data in period 1 and period 2 (empty and
filled circles, respectively). Both  are
one-zone SSC models (dashed line for period 1 and solid line for
period 2).}
\end{center}
\end{figure}

\section*{Acknowledgements}
AGILE is a mission of ASI, with co-participation of INAF and INFN. This work
was partially supported by ASI grants I/R/045/04, I/089/06/0, I/011/07/0 and by
the Italian Ministry of University and Research (PRIN 2005025417), (ASDC)
I/024/05/1. The MAGIC collaboration thanks the Instituto de Astrofisica de
Canarias for the excellent working conditions at the Observatorio del Roque de
Los Muchachos in La Palma and acknowledges support by the German BMBF and MPG,
the Italian INFN and Spanish MCINN, by ETH Research Grant TH 34/043, by the
Polish MNiSzW Grant N N203 390834, and the YIP of the Helmholtz Gemeinschaft.
The VERITAS collaboration is supported by grants from the U.S.  Department of
Energy, the National Science Foundation, and the Smithsonian Institution, by
NSERC in Canada, Science Foundation Ireland, and PPARC in the UK and
acknowledges the technical support staff at the FLWO. We also acknowledge the
Swift Team for carrying out the ToO observation. RMW's research is supported in
part by the DFG Cluster of Excellence ``Origin and Structure of the Universe''.

\begin{table}
\caption{SSC model parameters.}
\centering
\begin{tabular}{ccc}
\hline
  parameter      &    period 1     & period 2 \\
\hline   
 $\gamma_b$      & $3.6 \cdot 10^5  $  & $4.2 \cdot 10^5$ \\
 $\gamma_{max}$  & $1.3 \cdot 10^6$  & $1.3 \cdot 10^6$ \\
 $p_1$           & 2.22               & 2.1 \\
 $p_2$           & 4.5              & 5 \\
 $B$ [G]         & 0.1               & 0.1 \\
 $K$ [cm$^{-3}$] & $4 \cdot 10^{-4} $ & $6 \cdot 10^{-4} $ \\
 $\delta$        & 20                & 20 \\
 $\theta[^{\circ}]$  & 2                 &2 \\         
\hline
\end{tabular}
\end{table}

{}

\begin{thebibliography}{99}
\bibitem{lin92} Lin, Y. C. et  al. 1992, ApJ, \textbf{401},61
\bibitem{pun92} Punch, M. et al.  1992, Nature, \textbf{358}, 477
\bibitem{pad95} Padovani, P. \& Giommi, P. 1995, ApJ, \textbf{444}, 567
\bibitem{sam96} Sambruna, R.~M., {Maraschi}, L., {Urry}, C., 1996, ApJ, \textbf{463}, 444
\bibitem{fos98} Fossati, G. et al. 1998, MNRAS, \textbf{299}, 433
\bibitem{mar99} Maraschi, L. et al. 1999, ApJ, \textbf{526}, L81
\bibitem{fos08} Fossati, G. et al. 2008, ApJ, \textbf{677}, 906
\bibitem{wag08} Wagner, R.M., 2008, PoS(BLAZARS2008), \textbf{63}, 013 
\bibitem{ghi98} Ghisellini, G. et al. 1998, MNRAS, \textbf{301}, 451
\bibitem{aha00} Aharonian, F. A. 2000, NewA, \textbf{5}, 377
\bibitem{mue03} M\"{u}cke, A. et al. 2003, APh, \textbf{18}, 593
\bibitem{cos08} Costa, E. et al. 2008, ATEL $\# 1574$
\bibitem{pit08} Pittori, C. et al. 2008, ATEL $\#1583$
\bibitem{paper} Donnarumma, I. et al., 2009, ApJ, \textbf{691}, L13.
\bibitem{tav08} Tavani, M.  et al., 2008, Nucl. Instrum. Meth. A, \textbf{588}, 52
\bibitem{har99} Hartman, R. C., et al. 1999, ApJS, \textbf{123}, 79
\bibitem{loc95} Lockmann, F. J., \& Savage, B. D. 1995, ApJS, \textbf{97}, 1
\bibitem{mas04} Massaro, F. et al. 2004, A\&A, \textbf{422}, 103 
\bibitem{mas08} Massaro, F. et al. 2008, A\&A, \textbf{478}, 395
\bibitem{acc08} Acciari, V. A. et al. (VERITAS Collab.) 2008, ApJ, \textbf{679}, 1427
\bibitem{bai04} Baixeras, C. et al., 2004, Nucl. Instrum. Meth. A, \textbf{518}, 188
\bibitem{tes07} Tescaro, D. et al. (MAGIC Collaboration) 2008, in Proc. 30th ICRC, Merida, Mexico, Vol. 3, 1393 
\bibitem{dau97} Daum A. et al. (HEGRA Collab.) 1997, Astropart. Phys., \textbf{8}, 1
\bibitem{alb08} Albert et al. (MAGIC Collab.) 2008a, ApJ, \textbf{674}, 1037
\bibitem{dan07} Daniel, M. K. et al. 2007, in Proc. 30th ICRC, Merida, Mexico, Vol. 3, 283
\bibitem{hil85} Hillas A. M. 1985, in Proc. 19th ICRC, La Jolla, USA, Vol. 3, 445
\bibitem{hau01} Hauser, M. G., Dwek, E., 2001, ARA\&A, \textbf{39}, 249
\bibitem{rau08} Raue M., Mazin, D. 2008, Int. J. Mod. Phys. D, \textbf{17}, 1515
\bibitem{vil08} Villata, M., et al. 2008, A\&A,  \textbf{481}, L79
\bibitem{poo08} Poole, T. S. et al. 2008, MNRAS, \textbf{383}, 627
\bibitem{tav98} Tavecchio, F., et al. 1998, ApJ, \textbf{509}, 608 
\bibitem{vil99} Villata, M., \& Raiteri, C. M. 1999, A\&A, \textbf{347}, 30 
\bibitem{vil04} Villata, M., et al. 2004, A\&A \textbf{421}, 103 
\bibitem{lic08} Lichti, G. G., et al. 2008, A\&A,  \textbf{486}, 721
\bibitem{tra07} Tramacere, A., et al., 2007, A\&A, \textbf{467}, 501  



\end{thebibliography}
\end{document}